\documentclass[aps,prl,twocolumn,groupedaddress]{revtex4-1}
\usepackage{epstopdf}
\usepackage{graphicx}
\usepackage{subcaption}
\usepackage[font=small,labelfont=bf,
   justification=justified,
   format=plain]{caption}
\begin{document}


\title{Superconductivity in the dilute single band limit in reduced Strontium Titanate}


\author{Terence M. Bretz-Sullivan$^{[1]}$, Alexander Edelman$^{[2]}$, J. S. Jiang$^{[1]}$, Alexey Suslov$^{[3]}$, David Graf$^{[3]}$, Jianjie Zhang$^{[4]}$, Gensheng Wang$^{[4]}$, Clarence Chang$^{[2,4]}$, John E. Pearson$^{[1]}$, Alex B. Martinson$^{[1]}$, Peter B. Littlewood$^{[1,2]}$ and Anand Bhattacharya$^{[1]}$}

\affiliation{[1] Materials Science Division, Argonne National Laboratory, 9700 S. Cass Avenue, Lemont, IL 60439 \\ 
[2] Department of Physics, The University of Chicago, 5720 South Ellis Avenue, Chicago, IL 60637 \\
[3] The National High Magnetic Field Laboratory, 1800 E. Paul Dirac Drive, Tallahassee, FL 32310 \\
[4] High Energy Physics Division, Argonne National Laboratory, 9700 S. Cass Avenue, Lemont, IL 60439}


\date{\today}

\begin{abstract}

 We report on superconductivity in single crystals of SrTiO$_{3-\delta}$ with carrier densities \textit{n} $<  1.4 \times10^{18}cm^{-3}$, where only a single band is occupied. For all samples in this regime, the resistive transition occurs at  $T_{c} \approx 65 \pm 25 \ mK$.   We observe a zero resistance state for \textit{n} as low as $1.03 \times10^{17}cm^{-3}$, and a partial resistive transition for  \textit{n} $=  3.85 \times10^{16}cm^{-3}$. We observe low critical current densities, relatively high and isotropic upper critical fields, and an absence of diamagnetic screening in these samples. Our findings suggest an inhomogeneous superconducting state, embedded within a homogeneous high-mobility 3-dimensional electron gas. $T_{c}$ does not vary appreciably when \textit{n} changes by more than an order of magnitude, inconsistent with conventional superconductivity.

\end{abstract}

\pacs{}
\maketitle

The essential difficulties in understanding superconductivity in the low carrier density (\textit{n}) limit are captured in the BCS expression \cite{BCS} for the transition temperature $k_{B} T_{c} = 1.13\hbar \Theta_D \exp(-1/N(0) V)$. Here the Debye temperature $\Theta_D$ comes from a cutoff energy in the gap equation, set by knowledge of the characteristic scales of the microscopic mechanisms of electron attraction and repulsion, which is lacking outside the conventional phonon-mediated regime. $T_{c}$ is strongly reduced at low \textit{n}, unless the attractive pairing interaction strength $V$ compensates for the low density of states $N(0)$ at the Fermi level. Thus, the system does not necessarily flow to the usual BCS weak-coupling fixed point, opening up the possibility of pre-formed pairs and crossover to a Bose-Einstein condensate \cite{levy,eagles}. Similarly, the energy scales $\Omega$ of many possible pairing interactions are anti-adiabatic \cite{marel}, $\Omega >> E_F$, where pairing involving a retarded attractive interaction does not apply, Migdal's theorem is no longer valid, and perturbative calculations are questionable. 

In this light, superconductivity in doped SrTiO$_{3}$ (STO) has attracted much attention since the 1960's \cite{schooley, schooleySTO3-d}. It was the earliest example of a `dome' in the superconducting phase diagram of transition temperature $T_{c} \ vs. \ n$, for $6.9\times10^{18}cm^{-3} \textless \textit{ n} <  5.5\times10^{20}cm^{-3}$ \cite{koonce}, in a relatively dilute electron gas.  More recently, it was shown that superconductivity occurs in reduced STO at even lower densities with \textit{n} $< 1.5\times10^{18}cm^{-3}$, where only a single band is occupied \cite{lin1,lin2}.  Here superconductivity was found to persist down to $n=5.5 \times 10^{17} cm^{-3}$, with densities comparable to that of crystalline Bi, a superconductor with the lowest known carrier density of $n\approx3 \times 10^{17} cm^{-3}$ \cite{ramakrishnan} and $T_{c}\sim 0.5 \ mK$. 

Superconductivity in STO is derived from electrons pairing in three bands in the Ti 3d $t_{2g}$ manifold \cite{mattheiss}.  At low temperatures, the three degenerate bands are split at the $\Gamma$-point by a combination of spin-orbit interactions and octahedral rotations that lead to a tetragonal distortion.  Upon increasing the Fermi level through doping via chemical substitution or by introducing oxygen vacancies, electrons sequentially populate these three bands.  Despite the low carrier densities, for $n > 1.0\times10^{19}cm^{-3}$ tunneling and microwave measurements reveal an ordinary single-gap superconductor with a weak BCS coupling constant, smaller than that of aluminum, and a BCS-like $\Delta/k_{B}T_c$ \cite{swartz,mannhart,binnig}, although the microscopic pairing mechanism remains hotly debated \cite{cohen64,koonce,takada, Balatsky,Gorkov,ruhman,wolfe}.  Furthermore, STO is an incipient ferroelectric at low temperatures with a static dielectric constant $\varepsilon > 20,000$ \cite{muller}. As a result, when doped it remains metallic down to very low carrier densities with relatively high mobilities in high quality bulk single crystals and thin films \cite{behniaScience15, leighton, bhattacharyaNcomm16,stemmerNmat,Hwang}.
\begin{figure*}[t]
	\centering
	\includegraphics[width=0.9\textwidth]{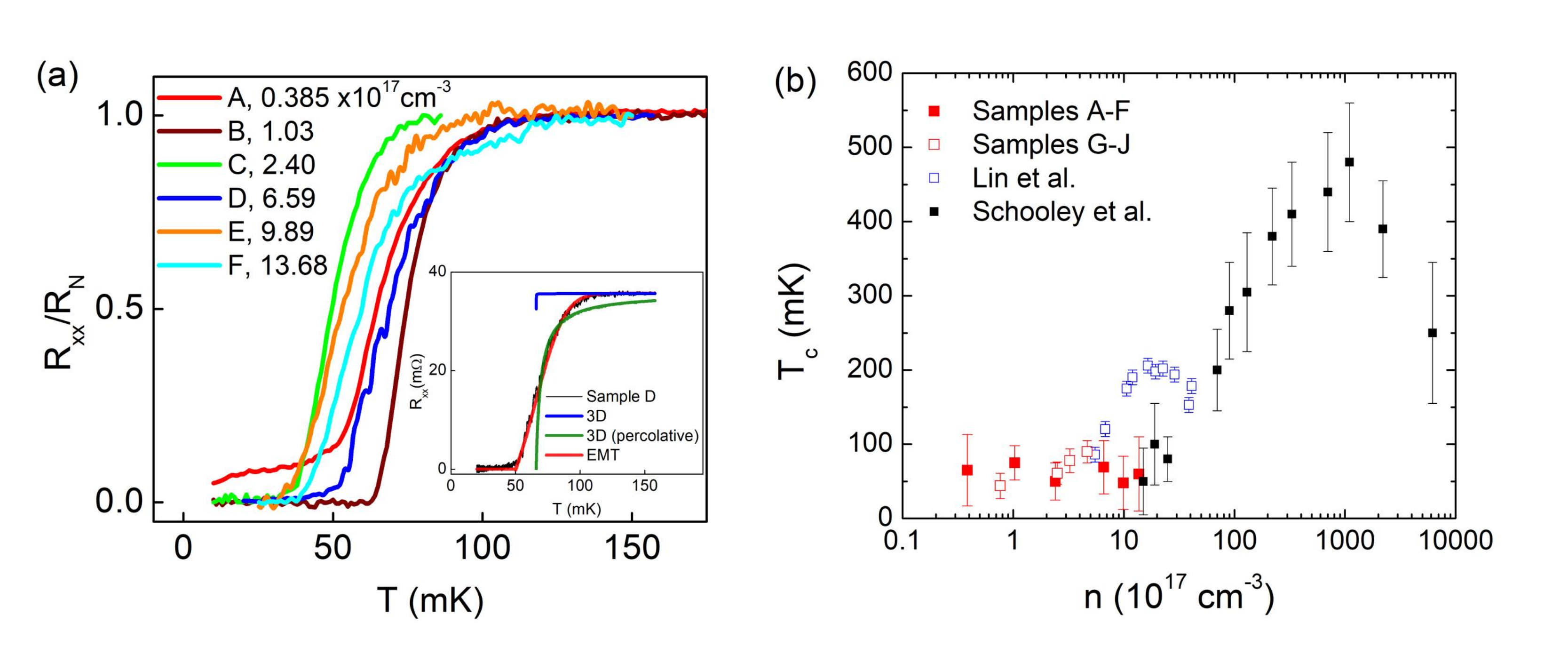}
	\caption{(a) Superconducting transitions of our samples labeled with increasing carrier density in units of $10^{17} cm^{-3}$, and (inset) paraconductivity and effective medium theory (EMT) model fits to sample \textit{D}. (b) The superconducting phase diagram, $T_{c} \ vs. \ n$, for reduced SrTiO$_{3-\delta}$ measured using resistive transitions of our samples \textit{A-F} (red solid squares) and \textit{G-J} (red open squares, measured along [011]) along with earlier work of Schooley \emph{et al.} (black) and Lin \emph{et al.} (blue)  superimposed \cite{schooleySTO3-d, lin1}.}
\end{figure*}

In this work, we report on superconductivity in SrTiO$_{3-\delta}$ in the dilute single band limit for $3.85\times10^{16}cm^{-3} < \textit{n} <  1.37 \times10^{18}cm^{-3}$, including densities much lower than previously reported \cite{lin1,schooleySTO3-d}. For all samples in this regime, we find  $T_{c} \approx 65 \pm 25 \ mK$.  Magnetotransport measurements reveal single frequency Shubnikov-de Haas (SdH) oscillations consistent with a single spherical Fermi surface.  Carrier densities inferred from SdH oscillations agree with those from Hall measurements, implying a homogeneous 3D electron gas. Furthermore, we find very low superconducting critical current densities, several orders of magnitude smaller than the depairing current for a homogenous condensate. This is accompanied by an absence of diamagnetic screening, and isotropic and substantially high upper critical fields.  Our data suggest that while the normal state carriers form a homogeneous 3-D electron gas, the superconducting state is inhomogeneous. Notably, in the single band limit, we find that $T_c$ does not vary appreciably when \textit{n} decreases by more than an order of magnitude, in sharp contrast with expectations from standard BCS theory.

All of our samples in this study show signatures of superconductivity with a high mobility metal in the normal state.  For sample preparation, and other details see Supplemental Material. In Fig. 1 (a), we plot $R_{xx}(T)$, normalized to the normal state resistance $R_{N}$, for 6 samples (\textit{A-F}) in zero magnetic field, while in Fig. 1 (b), we show a phase diagram of $T_{c} \ vs. \ n$ for all ten of our samples (solid/open red squares). For  samples \textit{G-J} (open red squares in Fig. 1 (b)), we measured transport properties along the [011] crystallographic direction, and have not applied magnetic fields below $1 \ K$ or measured superconducting critical currents (see Fig. S12). Eight of the samples exhibit a complete superconducting transition, with the highest onset temperature of $115 \ mK$ in sample \textit{F}.  Sample \textit{B} has the second largest $T_{c}$ of our set and nearly a fifth of the previous lowest carrier density at which superconductivity has been observed in reduced STO \cite{lin1}.  In contrast with Lin \emph{et al.} \cite{lin1} we do not observe a second `dome' near these densities. In fact,  $T_{c}\approx 65 \pm 25 \ mK$ for all of our samples, down to values of \textit{n} an order of magnitude lower than in the study of Lin \emph{et al.} \cite{lin1}. At higher \textit{n} our data agree with measurements of Schooley \emph{et al.} \cite{schooleySTO3-d} on the underdoped side of the main superconducting dome. Our data imply that for $n < 3\times10^{18} cm^{-3}$, $T_{c}$ does not vary much for nearly two orders of magnitude reduction in $n$. We note that recent theories for  superconductivity deep in the anti-adiabatic limit, where $n \rightarrow 0$,  predict that $T_{c}$ becomes independent of  $n$ \cite{kedem,fernandes}.  

At $n = 3.85\times10^{16}cm^{-3} $, sample \textit{A} exhibits an incomplete transition.  $R_{xx}(T)$ falls sharply below $T_{c}$ to $< 0.2\times R_{N}$, similar to sample \textit{D}, but then has a resistive tail below $50 \ mK$ going down to $\sim0.05\times R_{N}$ at $10 \ mK$.  This resistive transition and tail is reminiscent of behavior seen in ultrathin films of quenched deposited metals just on the threshold for global superconductivity \cite{jaeger,orr}, which were modeled as a network of Josephson coupled superconducting grains.  Global phase coherence occurs when the resistive dissipation between grains falls below $\hbar/e^{2}$ and is macroscopically tuned via $R_{N}$, which parameterizes static disorder \cite{jaeger,orr}. Possibly, sample \textit{A} with the lowest $n$ and highest $R_{N}$ values is just beyond the percolative limit for superconductivity in STO.

\begin{figure}[b]
	\centering
	\includegraphics[scale=0.3]{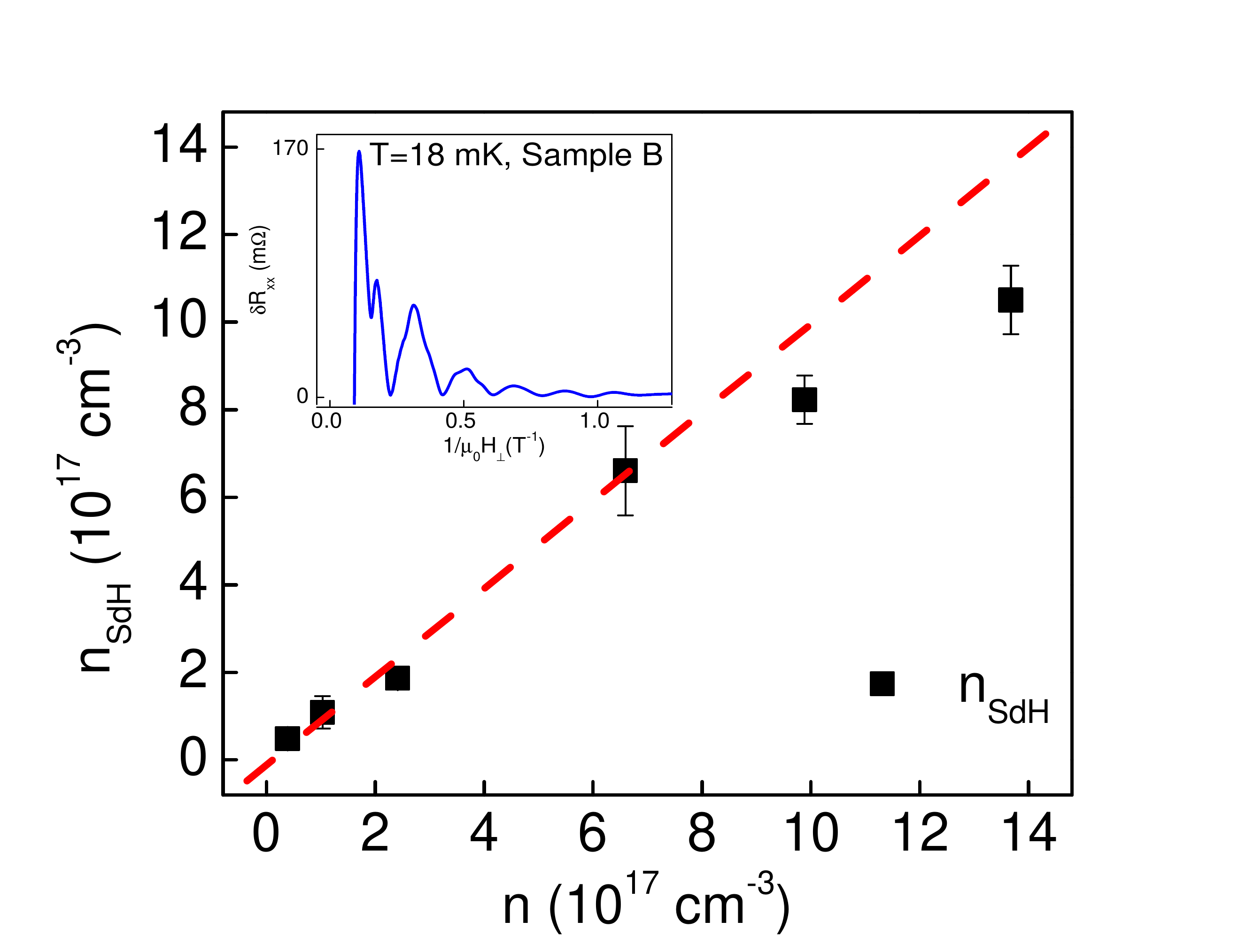}
	\caption{A plot of $n_{SdH} \ vs. \ n$ for samples \textit{A-F}.  The red dashed line is a guide for $n_{SdH}=n$.  Inset: Shubnikov-de Haas oscillations in sample \textit{B}.}
\end{figure}

In samples \textit{A-F}, where we have measured magnetoresistance in a dilution fridge, we observe Shubnikov-de Haas (SdH) oscillations in $R_{xx}(\mu_{0}H)$  periodic in $1/\mu_{0}H$, as each Landau level is swept through the Fermi energy (Fig. 2 inset and Supplemental Material). We extract the oscillations ($\delta R_{xx}$) by fitting the minima of the magnetoresistance traces with a fourth order polynomial in $\mu_{0}H$, and subtracting this polynomial fit from the raw data. The SdH oscillation period is related to the cross sectional area of the Fermi surface and thus the density of electrons in 3D by $\Delta(1/\mu_{0}H)=(\frac{16}{9\pi})^{1/3}\frac{n_{SdH}^{-2/3}}{\Phi_{0}}$.  Here $\Phi_{0}=\pi\hbar/e$ is the flux quantum, and we assume a spherical Fermi surface.  The single frequency observed in SdH oscillations is consistent with a single, small electron pocket contributing to transport, where the oscillation at the highest fields show evidence of spin splitting \cite{JAllen}. 

Additionally, in Fig. 2, we compare the measured Hall densities $n$ to $n_{SdH}$ to determine the homogeneity of the 3D electron gas underlying the superconducting state. $n$ is inferred by assuming a homogeneous electron gas through the thickness of the sample.  At larger carrier densities, $n_{SdH}$ deviates from $n$ as the onset of filling a second band begins \cite{lin1,lin2}. We note that in earlier measurements on a sample with a lower carrier density $n=1.05\times10^{17} cm^{-3}$, the SdH oscillations for magnetic fields applied in-plane and out-of-plane of the sample had nearly identical frequencies \cite{bhattacharyaNcomm16}, consistent with a 3-D Fermi surface symmetric about the principal axes.  

\begin{figure}[t]
	\centering
	\includegraphics[scale=0.3]{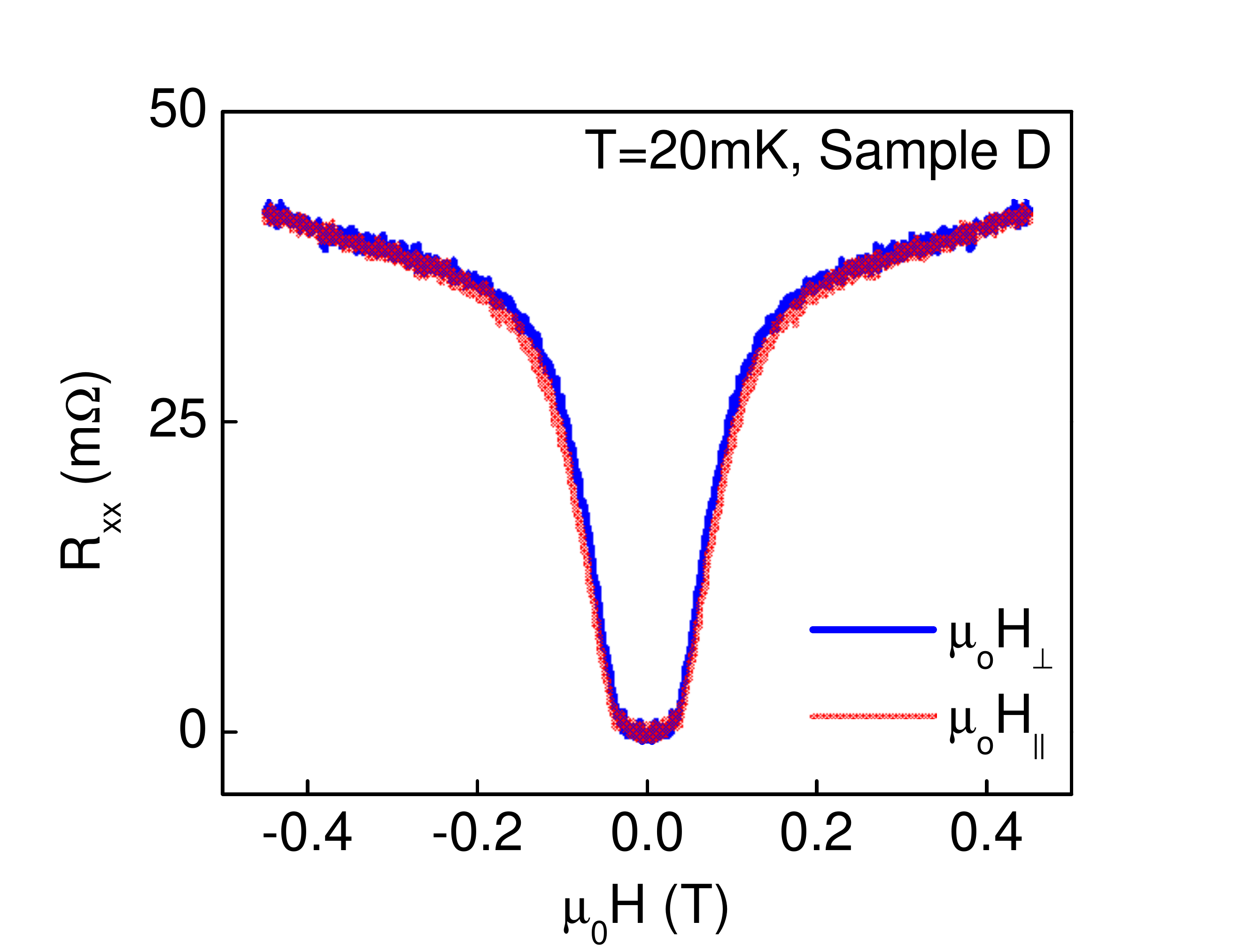}
	\caption{Low field $R_{xx}(T=20 \ mK,\mu_{0}H)$ for sample \textit{D} which exhibits an isotropic upper critical field ($\mu_{0}H_{c2}$).}  
\end{figure}

Next, we examine the upper critical field ($\mu_{0}H_{c2}$) of our samples in the low field magnetoresistance, where we define $\mu_{0}H_{c2}$ at $R_{N}/2$. As shown for sample \textit{D} (Fig. 3), the $R_{xx}(\mu_{0}H)$ sweeps are nearly the same for magnetic field orientations either out-of-the-plane (blue trace - $\mu_{0}H_{\bot}$) or in-plane and parallel with the applied current (red trace - $\mu_{0}H_{\parallel}$) (for other samples, see Supplemental Material).  We completely suppress superconductivity at $\mu_{0}H=0.2 \ T$.  For $\mu_{0}H > 0.2 \ T$, the finite slope at higher magnetic field values is due to the positive magnetoresistance of our samples in their normal state. If superconductivity in the very low $n$ limit were an artifact due to a 2D layer of oxygen vacancies in a near-surface region of the order of the London penetration depth $\lambda$ or thinner, $\mu_{0}H_{c2}$ would be enhanced in-plane relative to the out-of-plane value due to the absence of diamagnetic screening \cite{tinkham}.  Since this is not the case, our data are consistent with superconductivity permeating the bulk of the sample, and not confined to $\sim\lambda$ from the surface.  A simple estimate of $\lambda$, using London theory, would suggest 7 $\mu m < \lambda < $ 33 $\mu m $ (Supplemental Material).  However, as we state later, measuring $\lambda$ of these samples is not possible since they do not screen magnetic fields.
 
 The measured values of $\mu_{0}H_{c2}$ (Fig. 4 (b)) in our samples are found to increase with \textit{n}. Furthermore, barring the lowest doped sample, our $\mu_{0}H_{c2}$ values are significantly larger than expected from the corresponding $T_{c}$ values. They are also significantly larger than that measured in optimally doped Nb-STO ($n=2.6\times 10^{20} \ cm^{-3}$) where $\mu_{0}H_{c2} = 0.048 \ T$ for $T_{c}\sim 375\ mK$ \cite{collignon}.  We note that prior measurements on underdoped reduced STO samples, with carrier densities in the $\textit{n} = 10^{18} - 10^{19}\ cm^{-3}$ range ($100\ mK  < T_{c} <  230\ mK$) \cite{jourdan} also find larger values of $\mu_{0}H_{c2} \approx 0.3 - 0.4 \ T$. From the $\mu_{0}H_{c2}$ values in our samples, we estimate $\Delta/k_{B}T_c \sim 2-17$, significantly larger than the BCS value (Supplemental Material).

\begin{figure}[t]
	\centering	
 	\includegraphics[scale=0.4]{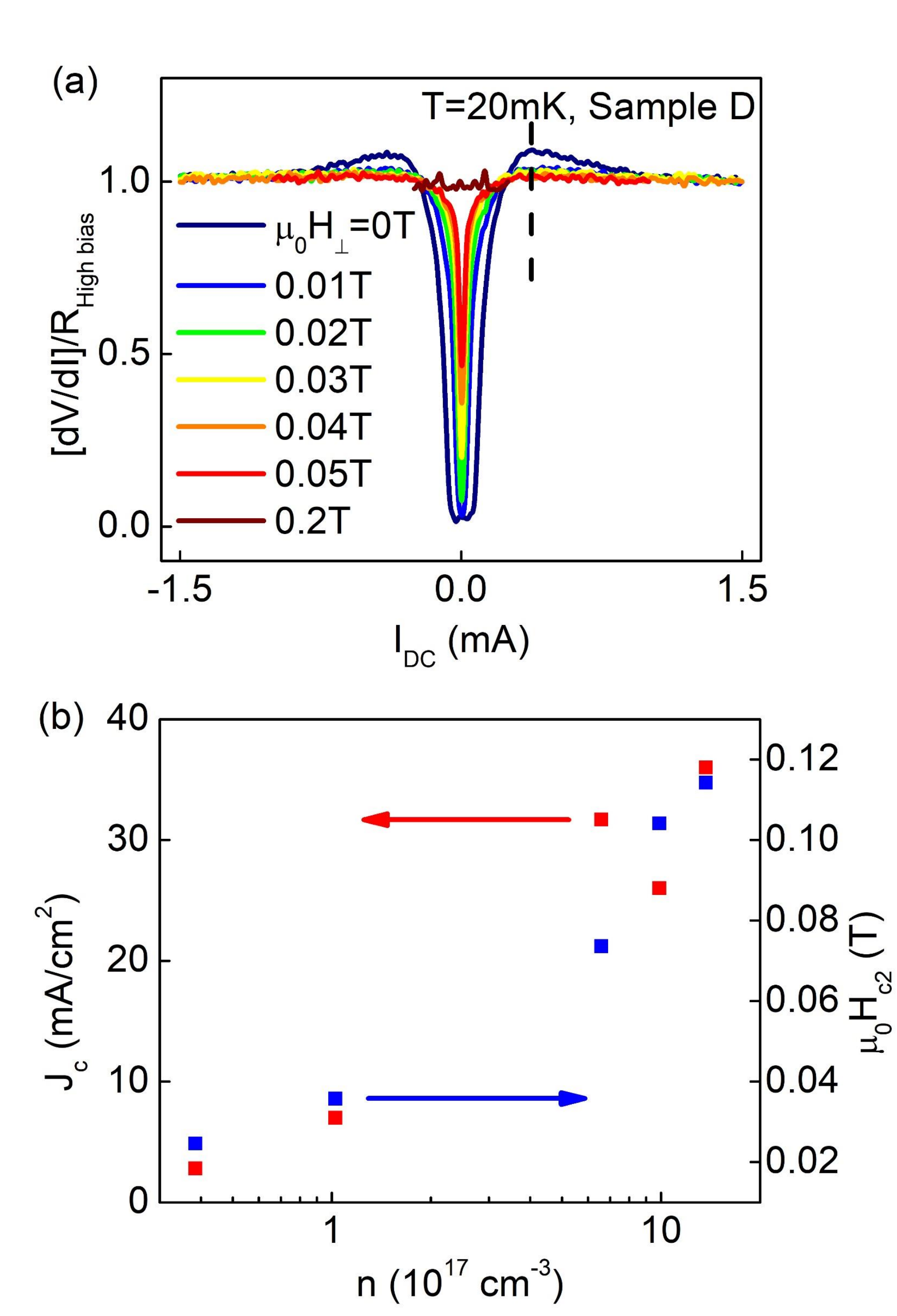}
 	 \caption{$J_{c}$ and $\mu_{0}H_{c2}$ measurements and results in the single band limit. (a) $dV/dI \ vs. \ I_{DC}$ measurements in sample \textit{D} with the suppression of $I_{c}$ as a function of magnetic field. The vertical dashed line corresponds to $I_{c}$. Each trace is normalized to 1 by the value of $dV/dI$ at $I_{DC,max}$, $R_{High \ bias}$. (b) Low $J_{c}$ and large $\mu_{0}H_{c2} \ vs.\  n$: $J_{c}$ falls considerably at lowest carrier densities.}
\end{figure}

A key finding of our work is the anomalously low critical current density ($J_{c}$) as seen in Fig. 4 (a,b).  We determine $I_{c}$ from the maxima in $dV/dI \ vs. \ I_{DC}$, denoted by the vertical black dashed line, and compute $J_{c}$ assuming that the supercurrent is uniformly distributed across the sample. We find $2.8\times 10^{-3} A/cm^{2} < J_{c}< 36\times 10^{-3} A/cm^{2}$.  The magnitude of $J_{c}$ is less than the estimated depairing current density by $10^{4}$ (Supplemental Material). The rapid suppression of $J_{c}$ at lowest doping, with relatively small variations in $T_{c}$, is suggestive of the onset of a percolative threshold with reduced carrier density, below which regions of similar $T_{c}$ fail to form a connected cluster that spans the measurement leads.   Analogous to the magnetoresistance measurements in Fig. 3, we suppress $I_{c}$ completely and isotropically at $\mu_{0}H= 0.2 \ T$. We note that  $J_{c}\sim10^{2} A/cm^{2}$ have been reported in reduced STO single crystals in earlier work \cite{jourdan}. However, these values of $J_{c}$ are higher than that in our highest doped samples by $\sim 3 \times 10^4$ while the carrier density estimated from  $T_{c}$ values in these samples is only $\sim 10$ times higher. This suggests that our samples are closer to a percolative threshold for superconductivity.

Another anomaly in our data is the width of the superconducting transition, $\Delta T_{c}$. In 3D, the excess conductivity due to superconducting fluctuations in a homogeneous superconductor above $T_{c}$, $\sigma^\prime \propto \epsilon^{-1/2}/\xi$, is a universal function of the reduced temperature $\epsilon = (T-T_{c})/T_{c}$ scaled by the superconducting coherence length $\xi$ \cite{skopcol,AL}. As shown in Fig. 1 (a) inset, this expression grossly underestimates the width of the transition. To incorporate the effects of a possible percolative state, we follow Char and Kapitulnik \cite{CharKapitulnik} in considering a static percolating network of fixed $T_{c}$ embedded in a bulk sample that remains normal. Here, superconducting fluctuations probe the underlying physical dimension of the system down to a percolative length scale $\xi_p$, below which the fractal dimension of percolation $d_{p}=4/3$ controls the paraconductivity, producing $\sigma^\prime_{\text{AL}} \propto \epsilon^{-4/3}$. Although the parameter values that best fit the transition are physically reasonable, i.e. $\xi_{p} \sim 10\xi$, we do not obtain qualitative agreement with the data. In general, the measured paraconductivity decays more quickly at high temperatures than any power-law fluctuation theory, whereas ordinarily $\Delta T_{c} \sim T_{c}$ would enhance thermal fluctuations. 

Lastly, we use a model in which the $T_{c}$ of sites is drawn from a Gaussian distribution of width $\sigma$.  Using an effective medium approximation (EMT) \cite{popcevic, nakamura} and a suitably chosen $\Delta T_{c}$ and $T_{c}$, we are able to reproduce the shape of the resistive transition, although some samples display slight excess conductivity at intermediate temperatures. In our samples $T_{c}$ and $\sigma$ vary unsystematically, and weakly, with  doping. The fact that each doping realization seems to produce a different macroscopic $T_c$ is difficult to explain away as a finite-size effect, due to the absence of resistive jumps in the data. The better fit obtained in the EMT than the fluctuation theory favors a picture in which different regions of the sample superconduct at different temperatures, with the bulk transition occurring when a fully superconducting pathway has been established across the sample. This is in contrast to fluctuations along a static percolative path that becomes superconducting at a single temperature. The unexplained residual conductance in some of the samples leaves open some role for fluctuations in a reduced dimension, which would be consistent with the observed low critical currents.  For details, see Supplemental Material.

At the lower doping levels in STO (\textit{n} $< 6.9\times10^{18}cm^{-3}$), there has been no evidence for Meissner screening \cite{schooleySTO3-d,cohen64,collignon}, a pre-requisite for establishing a homogeneous superconducting state. We attempted to measure changes in $\lambda$  of an unpatterned piece of sample \textit{E} using a tunnel diode oscillator setup down to $20 \ mK$. We were unable to resolve any signatures of a bulk superconducting transition (Supplemental Material), while we were able to do so in a LuPdBi sample, another low carrier density superconductor, using the same setup \cite{nakajima}.  Thus, to the best of our knowledge our samples do not show any diamagnetic screening in the superconducting state. Our result is consistent with earlier measurements where diamagnetic screening could not be observed in superconducting SrTiO$_{3-\delta}$ \cite{koonce}  for $\textit{n} < 6.9 \times 10^{18} cm^{-3}$, and in Nb-STO for $\textit{n} < 4.1 \times 10^{19} cm^{-3}$ \cite{collignon}.  

In conclusion, we observe superconductivity in SrTiO$_{3-\delta}$ single crystals in the single band limit, at carrier densities lower than reported in any material \cite{lin1,lin2,ramakrishnan}. The agreement between $\textit{n}$ and $\textit{n}_{SdH}$  establish that the normal state is a nearly homogeneous 3-D electron gas.  The lack of diamagnetic screening below $T_{c}$ suggests that the superconductivity is confined to regions smaller than $\lambda$ in at least one dimension.  Nonetheless, the relatively high values of $\mu_{0}H_{c2}$ indicate a well-developed superconducting gap $\Delta$, which increases with $\textit{n}$. This implies that superfluid density is inhomogeneously distributed in regions of $  \xi < \textit{l}_{sc} < \lambda$ that permeate the sample. However, $T_{c}$ remains nearly constant for all values of \textit{n} in the single band limit, suggesting an unconventional superconducting state. Our data imply a percolative transition as different such regions go superconducting, with $I_{c}$ being set by the weakest links, which depend on $n$.   Real space imaging of partial diamagnetic screening from our samples  may provide valuable clues about the spatial nature of this unusual superconducting state \cite{moler,moler2}.

All work at Argonne (sample preparation, transport and structural characterization, theoretical analysis) were supported by the U.S. Department of Energy, Office of Science, Basic Energy Sciences, Materials Science and Engineering Division.  We acknowledge the contributions from Gensheng Wang, Clarence Chang and Jianjie Zhang in assisting with low temperature transport measurements. Use of the Center for Nanoscale Materials, an Office of Science user facility, was supported by the U.S. Department of Energy, Office of Science, Office of Basic Energy Sciences, under Contract No. DE-AC02-06CH11357.  Dilution fridge based transport and tunnel diode oscillator measurements in magnetic fields were performed at the National High Magnetic Field Laboratory, which is supported by National Science Foundation Cooperative Agreement No. DMR-1157490 and DMR-1644779 and the State of Florida.  We also thank Carley Paulsen at the Institut N\'{e}el for his dilution fridge based SQUID magnetometry measurements.

\end{document}